\input harvmac
\overfullrule=0pt

\def\frac#1#2{{#1\over #2}}

\def\half{\frac12}

\def\ie{{\it i.e.}}
\def\eg{{\it e.g.}}



\Title{\vbox{\baselineskip12pt \hbox{}
}}
{\vbox{
\centerline{Lagrange Multipliers and Couplings}
\vskip 10pt
\centerline{in Supersymmetric Field Theory}
}}
\medskip

\centerline{\it
David Kutasov${}^1$
and Adam Schwimmer${}^2$}
\bigskip
\centerline{${}^1$EFI and Department of Physics, University of Chicago}
\centerline{5640 S. Ellis Av, Chicago, IL 60637, USA}
\centerline{${}^2$Department of Physics of Complex Systems,}
\centerline{Weizmann Institute of Science, Rehovot 76100, Israel}

\smallskip

\vglue .3cm
\bigskip
\noindent
In hep-th/0312098 it was argued that by extending 
the ``$a$-maximization'' of hep-th/0304128 away 
from fixed points of the renormalization group, 
one can compute the anomalous dimensions of chiral
superfields along the flow, and obtain a better
understanding of the irreversibility of RG flow
in four dimensional supersymmetric field theory. 
According to this proposal, the role of the running 
couplings is played by certain Lagrange multipliers 
that are introduced in the construction. We show 
that one can choose a parametrization of the space 
of couplings in which the Lagrange multipliers can 
indeed be identified with the couplings, and discuss 
the consequences of this for weakly coupled gauge theory.

\Date{8/2004}

\vfil\eject


\lref\ZamolodchikovGT{
A.~B.~Zamolodchikov,
``'Irreversibility' Of The Flux Of The Renormalization Group In A 2-D Field Theory,''
JETP Lett.\  {\bf 43}, 730 (1986)
[Pisma Zh.\ Eksp.\ Teor.\ Fiz.\  {\bf 43}, 565 (1986)].
}

\lref\KutasovSV{
D.~Kutasov and N.~Seiberg,
``Number Of Degrees Of Freedom, Density Of States And Tachyons In String Theory And Cft,''
Nucl.\ Phys.\ B {\bf 358}, 600 (1991).
}

\lref\CardyCW{
J.~L.~Cardy,
``Is There A C Theorem In Four-Dimensions?,''
Phys.\ Lett.\ B {\bf 215}, 749 (1988).
}

\lref\JackEB{
I.~Jack and H.~Osborn,
``Analogs For The C Theorem For Four-Dimensional Renormalizable Field Theories,''
Nucl.\ Phys.\ B {\bf 343}, 647 (1990).
}

\lref\CappelliYC{
A.~Cappelli, D.~Friedan and J.~I.~Latorre,
``C Theorem And Spectral Representation,''
Nucl.\ Phys.\ B {\bf 352}, 616 (1991).
}

\lref\CappelliKE{
A.~Cappelli, J.~I.~Latorre and X.~Vilasis-Cardona,
``Renormalization group patterns and C theorem in more than two-dimensions,''
Nucl.\ Phys.\ B {\bf 376}, 510 (1992)
[arXiv:hep-th/9109041].
}

\lref\BastianelliVV{
F.~Bastianelli,
``Tests for C-theorems in 4D,''
Phys.\ Lett.\ B {\bf 369}, 249 (1996)
[arXiv:hep-th/9511065].
}

\lref\ForteDX{
S.~Forte and J.~I.~Latorre,
``A proof of the irreversibility of renormalization group flows in four  dimensions,''
Nucl.\ Phys.\ B {\bf 535}, 709 (1998)
[arXiv:hep-th/9805015].
}

\lref\AnselmiYS{
D.~Anselmi, J.~Erlich, D.~Z.~Freedman and A.~A.~Johansen,
``Positivity constraints on anomalies in supersymmetric gauge theories,''
Phys.\ Rev.\ D {\bf 57}, 7570 (1998)
[arXiv:hep-th/9711035].
}

\lref\AnselmiAM{
D.~Anselmi, D.~Z.~Freedman, M.~T.~Grisaru and A.~A.~Johansen,
``Nonperturbative formulas for central functions of supersymmetric gauge  theories,''
Nucl.\ Phys.\ B {\bf 526}, 543 (1998)
[arXiv:hep-th/9708042].
}

\lref\AnselmiUK{
D.~Anselmi,
``Quantum irreversibility in arbitrary dimension,''
Nucl.\ Phys.\ B {\bf 567}, 331 (2000)
[arXiv:hep-th/9905005].
}

\lref\CappelliDV{
A.~Cappelli, G.~D'Appollonio, R.~Guida and N.~Magnoli,
``On the c-theorem in more than two dimensions,''
arXiv:hep-th/0009119.
}

\lref\SeibergPQ{
N.~Seiberg,
``Electric - magnetic duality in supersymmetric nonAbelian gauge theories,''
Nucl.\ Phys.\ B {\bf 435}, 129 (1995)
[arXiv:hep-th/9411149].
}

\lref\IntriligatorJJ{
K.~Intriligator and B.~Wecht,
``The exact superconformal R-symmetry maximizes a,''
Nucl.\ Phys.\ B {\bf 667}, 183 (2003)
[arXiv:hep-th/0304128].
}

\lref\KutasovIY{
D.~Kutasov, A.~Parnachev and D.~A.~Sahakyan,
``Central charges and U(1)R symmetries in N = 1 super Yang-Mills,''
JHEP {\bf 0311}, 013 (2003)
[arXiv:hep-th/0308071].
}

\lref\KutasovUX{
D.~Kutasov,
``New results on the 'a-theorem' in four dimensional supersymmetric field
theory,''
arXiv:hep-th/0312098.
}

\lref\IntriligatorMI{
K.~Intriligator and B.~Wecht,
``RG fixed points and flows in SQCD with adjoints,''
arXiv:hep-th/0309201.
}

\lref\CsakiUJ{
C.~Csaki, P.~Meade and J.~Terning,
``A mixed phase of SUSY gauge theories from a-maximization,''
JHEP {\bf 0404}, 040 (2004)
[arXiv:hep-th/0403062].
}

\lref\BarnesJJ{
E.~Barnes, K.~Intriligator, B.~Wecht and J.~Wright,
``Evidence for the Strongest Version of the 4d a-Theorem, via a-Maximization
Along RG Flows,''
arXiv:hep-th/0408156.
}

\lref\KutasovVE{
D.~Kutasov,
``A Comment on duality in N=1 supersymmetric nonAbelian gauge theories,''
Phys.\ Lett.\ B {\bf 351}, 230 (1995)
[arXiv:hep-th/9503086].
}

\lref\KutasovNP{
D.~Kutasov and A.~Schwimmer,
``On duality in supersymmetric Yang-Mills theory,''
Phys.\ Lett.\ B {\bf 354}, 315 (1995)
[arXiv:hep-th/9505004].
}

\lref\KutasovSS{
D.~Kutasov, A.~Schwimmer and N.~Seiberg,
``Chiral Rings, Singularity Theory and Electric-Magnetic Duality,''
Nucl.\ Phys.\ B {\bf 459}, 455 (1996)
[arXiv:hep-th/9510222].
}

\lref\BrodieVX{
J.~H.~Brodie,
``Duality in supersymmetric SU(N/c) gauge theory with two adjoint chiral  superfields,''
Nucl.\ Phys.\ B {\bf 478}, 123 (1996)
[arXiv:hep-th/9605232].
}

\lref\NovikovUC{
V.~A.~Novikov, M.~A.~Shifman, A.~I.~Vainshtein and V.~I.~Zakharov,
``Exact Gell-Mann-Low Function Of Supersymmetric Yang-Mills Theories From Instanton Calculus,''
Nucl.\ Phys.\ B {\bf 229}, 381 (1983).
}

\lref\ShifmanZI{
M.~A.~Shifman and A.~I.~Vainshtein,
``Solution Of The Anomaly Puzzle In Susy Gauge Theories And The Wilson Operator Expansion,''
Nucl.\ Phys.\ B {\bf 277}, 456 (1986)
[Sov.\ Phys.\ JETP {\bf 64}, 428 (1986\ ZETFA,91,723-744.1986)].
}

\lref\GrossCS{
D.~J.~Gross and F.~Wilczek,
``Asymptotically Free Gauge Theories. 2,''
Phys.\ Rev.\ D {\bf 9}, 980 (1974).
}

\lref\BanksNN{
T.~Banks and A.~Zaks,
``On The Phase Structure Of Vector - Like Gauge Theories With Massless Fermions,''
Nucl.\ Phys.\ B {\bf 196}, 189 (1982).
}

\lref\JackVG{
I.~Jack, D.~R.~T.~Jones and C.~G.~North,
``N = 1 supersymmetry and the three loop gauge beta function,''
Phys.\ Lett.\ B {\bf 386}, 138 (1996)
[arXiv:hep-ph/9606323].
}

\lref\JackCN{
I.~Jack, D.~R.~T.~Jones and C.~G.~North,
``Scheme dependence and the NSVZ beta-function,''
Nucl.\ Phys.\ B {\bf 486}, 479 (1997)
[arXiv:hep-ph/9609325].
}

\lref\LeighEP{
R.~G.~Leigh and M.~J.~Strassler,
``Exactly marginal operators and duality in 
four-dimensional N=1 supersymmetric gauge theory,''
Nucl.\ Phys.\ B {\bf 447}, 95 (1995)
[arXiv:hep-th/9503121].
}

\newsec{Introduction}

In the last year there was significant progress in the
study of four dimensional supersymmetric field theories. 
One interesting result was the demonstration \IntriligatorJJ\ 
that in an $N=1$ superconformal field theory, the R-current 
that appears in the superconformal multiplet (and thus can be
used to determine the scaling dimensions of chiral operators)
locally maximizes the combination of triangle anomalies
\eqn\aaa{a=3 {\rm tr} R^3-{\rm tr}R}
over the set of all possible R-currents. This observation
is especially useful in theories in which the infrared
R-current is preserved throughout the renormalization
group (RG) flow. As discussed in
\refs{\IntriligatorJJ\KutasovIY\IntriligatorMI\KutasovUX-\CsakiUJ},
this is often the case in four dimensional supersymmetric
field theories, and if it is, one can determine the IR $U(1)_R$
uniquely by using $a$-maximization. When the infrared R-current
involves accidental symmetries which only appear at the fixed
point, these techniques are less useful.\foot{Although a class of
accidental symmetries associated with the decoupling of gauge
invariant chiral superfields that reach the unitarity bound
\refs{\KutasovIY,\IntriligatorMI}, as well as some other cases
\KutasovUX, can still be treated this way.}

The work of \IntriligatorJJ\ also contributed to the quest 
for an analog of the c-theorem \ZamolodchikovGT\ in four
dimensional supersymmetric field theory. The central charge
$a$ \aaa\ is proportional to the coefficient of the Euler density
in the conformal anomaly of the theory on a curved spacetime
manifold \refs{\AnselmiAM,\AnselmiYS}. The latter
was conjectured many years ago to be an analog of the Virasoro
central charge in four dimensions \CardyCW. According to this
conjecture, $a$ should be always positive, and smaller at the
infrared fixed point of an RG flow than at the corresponding UV
fixed point. The fact that this quantity plays a role in the
solution of a different but related problem came as a pleasant surprise.

The authors of \IntriligatorJJ\ proposed an argument based on
$a$-maximization, that supports the validity of the conjectured
``$a$-theorem'', at least in cases where the infrared R-symmetry
is preserved throughout the RG flow. Since in these cases the
infrared $U(1)_R$ is a subgroup of the symmetry group of the
full theory, it is obtained by maximizing \aaa\ over a subset of
the R-currents that exist at the ultraviolet fixed point. Hence,
the corresponding maximum of $a$ should be lower.

An important loophole in this argument (which was pointed out
in \IntriligatorJJ) is that the maximum of \aaa\
that gives rise to the superconformal $U(1)_R$ is a local one. For
example, if the R-charges in the infrared are driven to values that
are significantly larger than those in the UV, $a$ \aaa\ can, in principle, 
grow in the process. This can only happen when the IR fixed point
is sufficiently far from the UV one (\ie\ non-perturbatively), but since
the techniques of \IntriligatorJJ\ are expected to be valid in a finite
region in coupling space
\refs{\IntriligatorJJ\KutasovIY\IntriligatorMI-\KutasovUX},
this possibility needs to be considered.
While gauge interactions  tend to reduce the R-charges, superpotentials
often have the opposite effect. Hence, in the presence of both gauge
interactions and superpotentials, the above argument needs to be sharpened.

This difficulty was overcome in \KutasovUX. The main idea was to extend
$a$-maximization away from fixed points of the renormalization group. It
was shown in \KutasovUX\ that one can define a function $a(\lambda_i)$
on a space of certain interpolating parameters $\lambda_i$, with the
following properties:
\item{(1)} The number of different interpolating parameters $\lambda_i$ is
the same as the number of independent couplings (gauge and superpotential)
in the Lagrangian of the theory in question. Each of the $\lambda_i$ is 
associated to a specific coupling. 
\item{(2)} The function $a(\lambda_i)$ smoothly interpolates between
the UV and IR fixed points of all the possible RG flows in the theory.
\item{(3)} There exist trajectories in the space of $\lambda_i$ along which
$a(\lambda_i)$ monotonically decreases as one goes from the UV to the IR.
Furthermore, it satisfies the gradient flow property 
\eqn\gradflow{\partial_i a=G_{ij}\beta^j~.}

\noindent
These results together with those of \IntriligatorJJ\ establish that
in all RG flows in which the superconformal $U(1)_R$ is a subgroup
of the full symmetry group of the theory, the $a$-theorem is satisfied.
They also lead to the natural question what is the precise relation 
between the interpolating parameters $\lambda_i$ which appear in the 
construction, and the gauge and superpotential couplings of the theory.

It was proposed in \KutasovUX\ that the $\lambda_i$ provide a
parametrization of the space of couplings of the quantum field 
theory. Evidence for this claim was provided by matching 
certain results in weakly coupled gauge theory with the analysis 
coming from $a$-maximization. However, the question whether
such an identification is fully consistent was left open. 

There are a number of reasons why the identification of
the interpolating parameters $\lambda_i$ with the field 
theory couplings  merits further investigation. First, the
$\lambda_i$ are introduced into the theory in a very different
way than the couplings, and it is not obvious apriori that the
two are related by a reparametrization. In the
usual description of a four dimensional field theory, the
dependence of the anomalous dimensions and the $\beta$ function
on the couplings is determined by performing loop calculations,
which rapidly become intractable as one goes to higher orders.
On the other hand, in \KutasovUX\ exact expressions were given
for all the anomalous dimensions as a function of $\lambda_i$,
without performing any loop calculations.\foot{One such
calculation is needed for each coupling, to normalize the
interpolating parameters $\lambda_i$.} It is natural to ask
what predictions these results provide for perturbative gauge
theory.

Also, it is usually said that the perturbative expansions
for quantities like anomalous dimensions and $\beta$-functions
in QFT have a vanishing radius of convergence. The radius of
convergence of the perturbative expansions in the $\lambda_i$
is finite, and it is interesting to see how that comes about.

The purpose of this note is to clarify the relation between 
the interpolating parameters $\lambda_i$ of \KutasovUX, and 
the field theory couplings. In section 2 we briefly review 
the construction of \KutasovUX, focusing for simplicity on 
the case of gauge theories with vanishing superpotential. In 
section 3 we show that the interpolating parameter $\lambda$ 
is related to the gauge coupling by an analytic transformation 
of the general form
\eqn\antrans{\lambda=\alpha+A_2\alpha^2+A_3\alpha^3+\cdots}
as proposed in \KutasovUX. In section 4 we discuss the
constraints placed by our discussion on the perturbative
expansion of anomalous dimensions in supersymmetric gauge
theory, and comment on other aspects of the analysis.
Some related results were independently found in \BarnesJJ. 

\newsec{$a$-maximization away from fixed points}

The setup for our discussion is an $N=1$ supersymmetric
gauge theory with gauge group $G$ and chiral superfields
$\Phi_i$ in the representations $r_i$ of the gauge group. 
The quadratic Casimir of $G$ in the representation $r$
will be denoted by $C_2(r)$. It is given in terms of the
generators of $G$ in that representation, $T^a$, by 
\eqn\Ccas{T^aT^a=C_2(r){\bf 1}_{|r|\times|r|}~.}
Here and below, $|r|$ denotes the dimension of the
representation $r$. Another useful object is $T(r)$ 
defined via the relation
\eqn\Tcas{{\rm Tr}_r (T^a T^b)=T(r)\delta^{ab}~.}
The two objects \Ccas, \Tcas\ are related by ($|G|$
is the dimension of the gauge group):
\eqn\quadcas{C_2(r)={|G|\over |r|} T(r)~.}
The gauge theory is asymptotically free when
\eqn\asfrcon{Q\equiv 3T(G)-\sum_i T(r_i)>0~,}
and we will restrict the discussion to this case.

At short distances, the gauge coupling $\alpha=g^2/4\pi$
goes to zero, and the theory approaches a free fixed point.
As the distance scale increases, the gauge coupling grows,
and the superfields $\Phi_i$ develop anomalous dimensions
$\gamma_i(\alpha)$. Our task is to understand this RG flow.

At the IR fixed point, the anomalous 
dimensions satisfy the constraint 
\eqn\fpcond{3T(G)-\sum_iT(r_i)(1-\gamma_i(\alpha^*))
=Q+\sum_iT(r_i)\gamma_i(\alpha^*)=0~,}
that follows from the NSVZ $\beta$-function 
\refs{\NovikovUC,\ShifmanZI}. This constraint has
a natural interpretation in terms of the R-charges of the
fields $\Phi_i$, $R_i$, which are related to the anomalous 
dimensions via the superconformal algebra:
\eqn\delrel{\Delta(\Phi_i)=1+{1\over2}\gamma_i={3\over2}R_i~.}
Plugging this into \fpcond\ we find 
\eqn\anomcond{T(G)+\sum_i T(r_i)(R_i(\alpha^*)-1)=0~,}
which can also be thought of as the condition for the R-symmetry
with $R(\Phi_i)=R_i(\alpha^*)$ to be anomaly free at
non-zero gauge coupling.

As we flow from short to long distances, the 
R-charges $R_i(\alpha)$ \delrel\ flow from 
$2/3$ in the UV to a solution of \anomcond\ 
in the IR. In general, this equation has many 
solutions. The one that gives the values of 
$R_i$ at the IR fixed point locally maximizes 
$a$ \aaa, subject to the constraint \anomcond\ 
\IntriligatorJJ. 

As was shown in \KutasovUX, a useful way of solving 
for $R_i(\alpha^*)$ is to take the constraint into 
account by means of a Lagrange multiplier:\foot{We 
have rescaled $\lambda$ by a factor $2|G|/\pi$ relative
to its normalization in \KutasovUX.}
\eqn\genera{ a(R_i,\lambda)=
2| G|+\sum_i |r_i|\left[ 3(R_i-1)^3-(R_i-1)\right]-
\lambda{2|G|\over\pi}\left[T(G)+\sum_i T(r_i)(R_i-1)\right]~.}
If we first locally maximize \genera\ with respect to the $R_i$,
for fixed $\lambda$, solve for $R_i(\lambda)$, and substitute
back into \genera, we get a central charge that depends only on
$\lambda$, $a(R_i(\lambda),\lambda)$ which has the following
properties \KutasovUX:
\item{(1)} At $\lambda=0$, one finds the R-charges and central 
charge \aaa\ of the free UV fixed point. 
\item{(2)} The infrared fixed point of the gauge theory
corresponds to a local minimum of $a$ at a positive value of
$\lambda$, $\lambda=\lambda^*$. $R_i(\lambda^*)$ are
the IR R-charges and $a(R_i(\lambda^*),\lambda^*)$ gives the 
value of the central charge \aaa\ at the infrared fixed point.
\item{(3)} As one varies $\lambda$ between $0$ and $\lambda^*$,
the central charge $a$ monotonically decreases. It satisfies 
the relation
\eqn\deralam{{da\over d\lambda}=- 
{2|G|\over\pi}\left[T(G)+\sum_i T(r_i)(R_i(\lambda)-1)\right]~.}
The right hand side is proportional to the 
$\beta$-function for the gauge coupling. 
\item{(4)} The R-charges along the flow are given by 
\eqn\rriill{R_i(\lambda)=1-{1\over3}\left[1+{2\lambda\over\pi}
C_2(r_i)\right]^{\half}~.}

\noindent
As emphasized in \KutasovUX, these properties strongly
suggest that $\lambda$ should be identified with the
gauge coupling in some scheme (or parametrization of the
space of gauge theories). We next show that this is indeed
the case.

\newsec{Lagrange multipliers and gauge couplings in 4d SYM theory}

In order to understand the relation between the gauge coupling
$\alpha$ and the Lagrange multiplier $\lambda$, we have to 
take a closer look at the freedom of reparametrizing the space
of theories.

Suppose we start in a scheme in which
the running of the gauge coupling is governed by the NSVZ
$\beta$-function \refs{\NovikovUC,\ShifmanZI}. The latter
can be written as
\eqn\betansvz{\beta(\alpha)=f(\alpha)
\left[Q+\sum_iT(r_i)\gamma_i(\alpha)\right]}
with
\eqn\ffaa{f(\alpha)=-{\alpha^2\over2\pi}
{1\over 1-{\alpha\over2\pi}T(G)}~.}
We will assume that throughout the RG flow,
the gauge coupling never becomes large enough
so that $f(\alpha)$, \ffaa, becomes singular. For
the purposes of the present discussion, this assumption
is not restrictive, since we will work in the
infinitesimal vicinity of $\alpha=0$ (to all orders in
$\alpha$), where the denominator of \ffaa\ is harmless.

Any other scheme will in general differ from
this one by a reparametrization of the coupling,
and by a coupling-dependent rescaling of the
chiral superfields:
\eqn\repalphi{\eqalign{
\alpha\to&\tilde\alpha(\alpha)\cr
\Phi_i\to&\tilde\Phi_i=\sqrt{F_i(\alpha)}\Phi_i~.\cr
}}
As usual, we would like all schemes to agree
at weak coupling, so that:
\eqn\smallalpha{\eqalign{
&\tilde\alpha=\alpha(1+a_1\alpha+a_2\alpha^2+\cdots)\cr
&F_i(\alpha)=1+b^{(1)}_i\alpha+b^{(2)}_i\alpha^2+\cdots .\cr
}}
The anomalous dimensions transform as follows under 
\repalphi, \smallalpha: 
\eqn\chgammai{\tilde\gamma_i(\tilde\alpha)=
\gamma_i(\alpha)-\beta(\alpha){d\over d\alpha}\log F_i(\alpha)~.}
A short calculation shows that after the transformation \repalphi,
\chgammai, the $\beta$-function takes the form
\eqn\newbeta{\tilde\beta(\tilde\alpha)=\tilde f(\tilde\alpha)
\left[Q+\sum_iT(r_i)\tilde\gamma_i(\tilde\alpha)\right]}
with 
\eqn\newff{\tilde f(\tilde\alpha)=
f(\alpha){d\tilde\alpha\over d\alpha}
{1\over 1-f(\alpha){d\over d\alpha}\sum_i T(r_i)\log F_i(\alpha)}~.}
Thus, we see that the fact that the $\beta$-function is proportional
to $Q+\sum_iT(r_i)\gamma_i(\alpha)$ is preserved under
reparametrizations. The only thing that changes is the prefactor
$f(\alpha)$ in \betansvz. This fact implies that as long 
as the prefactor $\tilde f(\tilde\alpha)$ in \newbeta\ is
non-singular along the RG flow, the infrared fixed point 
corresponds to a solution of \fpcond, in any parametrization 
of the gauge coupling and of the superfields $\Phi_i$.

Naively, one might be tempted at this point to argue as follows.
The freedom \smallalpha, \chgammai\ allows one to bring the
anomalous dimensions to a form linear in $\alpha$ (recall that
the $\beta$-function \betansvz\ starts like $\alpha^2$ for small $\alpha$),
\eqn\lingamma{\tilde\gamma_i(\tilde\alpha)=-{\tilde\alpha\over\pi}
C_2(r_i)~.}
To find the IR fixed point it is natural to set the term
in square brackets in \newbeta\ to zero,
\eqn\naivezero{Q+\sum_iT(r_i)\tilde\gamma_i(\alpha^*)=0~,}
as in \fpcond.

Substituting \lingamma\ into \naivezero, we find
\eqn\naiveansw{\eqalign{
\tilde\alpha^*=&{\pi Q\over \sum_iT(r_i)C_2(r_i)}\cr
\tilde\gamma_i(\tilde\alpha^*)=&-{Q C_2(r_i)\over
\sum_iT(r_i)C_2(r_i)}~.\cr}}
Interestingly, this answer is incorrect! Recall \KutasovUX\
that the correct answer is obtained by solving \anomcond,
\rriill; the resulting anomalous dimensions disagree with
\naivezero\ beyond leading order in $\lambda^*$ (or more 
precisely, in $Q$; see discussion in section 4).

What went wrong? Evidently, the prefactor
$\tilde f(\tilde\alpha)$ in \newbeta\ is not
harmless in this case. The transformation to
the coordinates in which \lingamma\ is valid
must have the property that it pushes the solution
of \naivezero, \naiveansw\ beyond the regime in
which \newff\ is well behaved. The lesson from
this is that when we make coordinate transformations
\repalphi, we have to make sure that the solution of
eq. \fpcond, \naivezero\ is not pushed outside of the
regime of validity of \newbeta.

Now, suppose the anomalous dimensions $\gamma_i(\alpha)$
have been calculated in some scheme (\eg\ that of
\refs{\JackVG,\JackCN}) with the $\beta$-function 
given by \betansvz\ (with 
$f(\alpha)=-{\alpha^2\over2\pi}(1+O(\alpha))$, 
not necessarily given exactly by \ffaa). Perform 
a reparametrization \smallalpha\ such that\foot{This
expression is equivalent to \rriill, using \delrel,
with $\lambda$ replaced by $\tilde\alpha$.}
\eqn\repgam{\tilde\gamma_i(\tilde\alpha)=1-
\left[1+{2\tilde\alpha\over\pi}C_2(r_i)\right]^{\half}~.}
By construction, after the reparametrization the
$\beta$-function \newbeta\ vanishes when the expression
in square brackets \naivezero\ is set to zero. Thus, it is
natural to expect the dangerous prefactor $\tilde f(\tilde\alpha)$
in \newbeta\ to be well behaved in this case, so that the
transformation $(\alpha,\Phi_i)\to(\tilde\alpha,\tilde\Phi_i)$
is non-singular throughout the RG flow for sufficiently
weak coupling. We will assume this to be the case and
will verify it in the next section, when we study the
theory perturbatively.

The assumption that $\tilde f(\tilde\alpha)$ is regular
throughout the RG flow implies that the map 
$\alpha\to \tilde\alpha(\alpha)$ is monotonic: 
as $\alpha$ increases from $0$ to $\alpha^*$,
$\tilde\alpha$ increases from $0$ to $\tilde\alpha^*$.
Any turning points of the map would give rise to zeroes
of $d\tilde\alpha\over d\alpha$, and thus of $\tilde f$
(the other factors in \newff\ can not cancel such zeroes).

To recapitulate, in this section we have presented an
explicit prescription for going from a general parametrization
of the space of $N=1$ SYM theories with matter, to the
one naturally given by the Lagrange multiplier construction
of \KutasovUX. Key elements are the fact that the NSVZ
$\beta$-function, \betansvz, transforms covariantly under
analytic reparametrizations \repalphi, and the fact that under
the specific reparametrization that leads to \repgam\ we expect
the prefactor in \newbeta\ to be analytic at weak coupling.

\newsec{Weakly coupled SYM theory}

As mentioned in the introduction, the construction
of \KutasovUX\ and the previous section gives rise
to predictions for the perturbative expansion of 
anomalous dimensions in gauge theory. In order to
see what those predictions are, we start with the
prefered scheme in which the anomalous dimensions 
are given by \repgam, and perform a general coordinate 
transformation \smallalpha, \chgammai. 

We find the following structure:
\eqn\expgam{\gamma_i(\alpha)=
1-\left[1+{2\tilde\alpha\over\pi}C_2(r_i)\right]^{\half}
+\beta(\alpha){F_i'(\alpha)\over F_i(\alpha)}~.}
Expanding
\eqn\gammaialpha{\gamma_i(\alpha)=\sum_{n=1}^\infty
\gamma_i^{(n)}\alpha^n~,}
the leading contribution to $\gamma_i^{(n)}$ comes
from replacing $\tilde\alpha$ by $\alpha$ in the
square root in \expgam, and expanding to $n$'th order. 
The second term in \expgam\ can be neglected at
this order, due to the overall factor of the 
$\beta$-function in it.\foot{This statement will 
be made more precise below.} 

The expansion of the square root gives
\eqn\expggiinn{\gamma_i^{(n)}=
\left[-{C_2(r_i)\over2\pi}\right]^n
{2(2n-2)!\over n!(n-1)!}+\cdots}
There are two types of corrections to \expggiinn.
One comes from substituting subleading terms in 
the expansion of $\tilde\alpha(\alpha)$ \smallalpha\ 
in \expgam. This completes \expggiinn\ to
\eqn\nloop{\gamma_i^{(n)}= a^{(n)}_n\left(C_2(r_i)\right)^n
+a^{(n)}_{n-1}\left(C_2(r_i)\right)^{n-1}+\cdots +
a^{(n)}_1C_2(r_i)~,}
where $a^{(n)}_n$ is given by \expggiinn, 
and the lower terms can be expressed in terms of
the expansion coefficients $a_i$ in \smallalpha. 

The second source of corrections comes from the expansion
of the last term in \expgam. It is sensitive to the precise
form of the prefactor $f(\alpha)$ in the $\beta$-function
\betansvz\ (which may or may not be given by \ffaa, depending
on the scheme), and the expansion coefficients 
$b_i^{(n)}$ in \smallalpha. 

The above predictions can be compared to  ``data.''
Results for the anomalous dimensions up to three loop
order appear in \refs{\JackVG,\JackCN}:
\eqn\threeloop{\eqalign{
\gamma_i(\alpha)=-&{\alpha\over\pi}C_2(r_i)+\cr
&{\alpha^2\over4\pi^2}\left[2\left(C_2(r_i)\right)^2-QC_2(r_i)\right]-\cr
&{\alpha^3\over8\pi^3}\left[4\left(C_2(r_i)\right)^3-Q\left(C_2(r_i)\right)^2+
AC_2(r_i)\right]+O(\alpha^4)\cr}}
The coefficient $A$ is also given in \refs{\JackVG,\JackCN}; 
we will not need its form here, and so do not quote it.

The first thing to note is that the leading term at each order
in $\alpha$ is in exact agreement with \expggiinn. The agreement
of the one and two loop coefficients was checked before in
\KutasovUX; the three loop check is new. This check was 
independently made in \BarnesJJ, where the prediction \expggiinn, 
\nloop\ was also derived, assuming the relation between the 
Lagrange multiplier of \KutasovUX\ and the gauge coupling.

Comparing the term that goes like $C_2(r_i)$ on the second line
of \threeloop\ we conclude that the relation between the
gauge couplings \smallalpha\ is
\eqn\firstcorr{\tilde\alpha=
\alpha\left(1+{Q\alpha\over4\pi}+O(\alpha^2)\right)}
Using this redefinition at order $\tilde\alpha^2$ in
\expgam, we find a term in $\gamma_i$ that goes like
$${Q\alpha^3\over4\pi^3}(C_2(r_i))^2~.$$
This is larger by a factor of two than the second term
on the third line of \threeloop. The difference can be
accounted for by using the last term in \expgam. One
finds:
\eqn\fipert{F_i(\alpha)=1+{\alpha^2\over8\pi^2} (C_2(r_i))^2
+O(\alpha^3)~.}
We see that to three loop order, the structure is as
anticipated in \KutasovUX\ and section 3. The leading
terms in $\gamma_i^{(n)}$ \nloop, which are scheme
independent, sum up to the square root \repgam. Using
the subleading terms one can determine the reparametrization
expansion \smallalpha\ perturbatively in $\alpha$. One finds
that, to this order, the coefficients $a_i$, $b_i^{(n)}$ in
\smallalpha\ are polynomials in the Casimirs. We expect this
to persist to all orders in $\alpha$.

Note that if we treat the quadratic Casimirs $C_2(r_i)$,
$T(r_i)$, $Q$, as formal variables (ignoring the fact that
they take discrete values), the problem has a scaling symmetry
in any renormalization scheme. Under this symmetry, the
quadratic Casimirs transform with weight one, $\alpha$ 
and the $\beta$-function have weight $-1$, and the anomalous
dimensions have weight zero. 

The above symmetry is respected by our Lagrange multiplier
inspired expression \repgam, and by any gauge theory loop
expansion, such as \threeloop. The expansion coefficients
$a_n$, $b_i^{(n)}$ in \smallalpha\ have weight $n$, while the
coefficients $a^{(n)}_i$ in \nloop\ have weight $n-i$. If no
inverse powers of Casimirs appear, they are polynomials of
the relevant degree in the Casimirs. As mentioned above, we
expect this to be the case to all orders in $\alpha$.

The notion of weakly coupled RG flows can be made precise
as follows. If we treat the Casimirs as free continuous parameters
(something that becomes precisely justified for large rank gauge
groups, but is often a very good approximation for finite rank as
well), it is well known that as $Q$ \asfrcon\ goes to zero the theory
loses asymptotic freedom, and if $Q$ is positive and small,
the IR fixed point is weakly coupled and can be
treated perturbatively \refs{\GrossCS,\BanksNN}. As one flows
from the UV to the IR in these situations, $\alpha$ varies between
zero and a value of order $Q$. Therefore, it makes sense to expand
the anomalous dimensions \gammaialpha\ and other quantities in a
double series in $\alpha$ and $Q$.

The expansion coefficients $a^{(n)}_i$ in \nloop\ are polynomials
of degree $n-i$ in $Q$. In particular, $a^{(n)}_n$ is a constant
(given by \expggiinn), $a^{(n)}_{n-1}$ is a first order polynomial
(examples of which appear in \threeloop), etc. Another way of thinking
about the construction of section 3 is that starting with a set of
$\gamma_i(\alpha)$ with the properties outlined above, we can perform
a transformation of the form \smallalpha, \chgammai\ to bring
the anomalous dimensions to a form where $Q$ does not appear, 
without introducing non-analiticity in $Q$ in the prefactor 
$\tilde f(\tilde\alpha)$ in \newff. One can then show that, 
up to a $Q$-independent reparametrization, the anomalous 
dimensions must be given by \repgam.

We finish with a few comments:

\item{(1)} We found that there is an analytic redefinition
of the coupling $\alpha$ and of the superfields $\Phi_i$, 
after which the anomalous dimensions are given by a
perturbative series with a finite radius of convergence
\repgam. It would be nice if this parametrization arose
from a natural renormalization scheme in perturbative SYM
theory. We do not know whether this is the case.
\item{(2)} We expect the coefficients $a_i$, $b_i^{(n)}$
\smallalpha, which determine the above redefinition, to be
polynomials in the Casimirs, whose rank was indicated earlier
in this section. This would guarantee the regularity of 
$\tilde f(\tilde\alpha)$ \newff\ to all orders in $Q$. 
We have not proven this property -- it would be very 
nice to do that.
\item{(3)} While we focused here on the case of a gauge theory
with a semi-simple gauge group and vanishing superpotential,
the discussion can be easily extended to situations with
multiple couplings, as in \KutasovUX. In order to achieve
a quantitative understanding of the flow of couplings with 
the scale in this case, one needs to know the metric governing 
the gradient flow of the central charge $a$, \gradflow.
Some perturbative results on this metric appear in
\refs{\KutasovUX,\BarnesJJ}.
\item{(4)} The most important open problem remains to 
extend the discussion beyond the regime of validity
of the NSVZ $\beta$-function, or more generally to the
regime where the description in terms of the original
degrees of freedom $W_\alpha$, $\Phi_i$ breaks down.

\vskip 1cm
\noindent{\bf Acknowledgments:}
We thank O. Aharony, M. Berkooz and S. Elitzur for discussions.
D.K. thanks the Weizmann Institute for hospitality, and the 
Einstein Center, the Center for Basic Interactions of the 
Israeli Academy of Science, and the DOE for partial support.

\listrefs
\bye